\documentclass[aps,prc,preprint,showpacs,preprintnumbers,amsmath,amssymb] {revtex4-1}
\usepackage{amsmath}
\usepackage{amssymb}
\usepackage{graphicx}
\usepackage{subfigure}
\usepackage{rotating}
\usepackage{natbib}
\usepackage{txfonts}
\usepackage{color}
\usepackage{bbold}
\def\beq{\begin{equation}}
\def\eeq{\end{equation}}
\def\bea{\begin{eqnarray}}
\def\eea{\end{eqnarray}}

\def\eqref#1{Eq.~(\ref{eq:#1})}

\def\be{\begin{equation}}
\def\ee{\end{equation}}
\def\bg{\begin{eqnarray}}
\def\en{\end{eqnarray}}

\long\def\Omit#1{}

\DeclareGraphicsExtensions{ .pdf, .jpg, .eps}
\begin{document}
\title{Charmed-baryon production in antiproton-proton collisions within an
effective Lagrangian model}
\author{R. Shyam}
\affiliation{Saha Institute of Nuclear Physics, 1/AF Bidhan Nagar, Kolkata 
700064, India {
\footnote {radhey.shyam@saha.ac.in}}}

\date{\today}
\begin{abstract}
We study the productions of charmed baryons ${\bar \Lambda}_c^- 
\Lambda_c^+$, ${\bar \Lambda}_c^- \Sigma_c^+$, and ${\bar \Sigma}^-
\Sigma_c^+$ in the antiproton-proton collisions within an effective 
Lagrangian model that has only the baryon-meson degrees of freedom and 
involves the physical hadron masses. The baryon production proceeds via 
the $t$-channel exchanges of $D^0$ and $D^{*0}$ mesons in the initial 
collision of the antiproton with the target proton. The distortion 
effects in the initial and final states are accounted for by using an 
eikonal approximation-based procedure. We find that the reaction 
amplitudes of all the production channels are dominated by the $D^{*0}$ 
meson-exchange diagrams. We discuss the relative roles of tensor and 
vector components of the $D^{*0}$ coupling in the $D^{*0}$ meson-exchange 
component of the total production cross sections. The magnitudes of the 
cross sections are predicted for each final state for the range of beam 
momenta of relevance to the ${\bar P}ANDA$ experiment.  
\end{abstract}
\maketitle

\section{Introduction}

The current experimental information about the production of the ground 
state charmed baryons has been derived mostly from the electron-positron 
annihilation experiments (see, e.g. Refs.~\cite{dub03,zie11,kat14}). In 
the near future, charmed-baryon production will be studied in the 
antiproton-proton (${\bar p}p$) annihilation using the "antiproton 
annihilation at Darmstadt" (${\bar P}ANDA$) experiment at the Facility 
for Antiproton and Ion Research (FAIR) in GSI, Darmstadt (see, e.g., 
Ref.~\cite{wie11}). The advantage of using antiprotons in the study of 
charmed baryons is that in ${\bar p}p$ collisions the production of extra 
particles is not needed for the charm conservation, which reduces the 
threshold energy as compared to, say, $pp$ collisions. The beam momenta 
of antiprotons in this experiment will be well above the thresholds of 
the productions of ${\bar \Lambda}_c^- \Lambda_c^+$ and ${\bar \Sigma}^- 
\Sigma_c^+$ charmed baryons in the ${\bar p}p$ collisions. For the 
planning of this experiment, reliable theoretical estimates of the cross 
sections of these reactions would be of crucial importance. The
production rates of these reactions are also the key requirement in the 
implementation of other programs~\cite{ern09} of the ${\bar P}ANDA$ 
experiment, {\it e.g.}, spectroscopy of charmed hadrons (see, e.g., Ref.
\cite{wie11}), production of charm hypernuclei \cite{dov77,shy17}, $D$ 
mesonic nuclei~\cite{tsu99,gar10,gar12,yam16}, and medium effects on the 
charmed-hadron properties~\cite{tsu03,tsu04,jim11,tol13,kre16,shy16}. 

The investigations of the production of heavy flavor hadrons provide an 
additional handle for the understanding of quantum chromodynamics, 
the fundamental theory of the strong interaction. The presence of the 
heavy charm quark along with the light quark(s) leads to two energy 
scales in such systems. This allows the construction of an effective 
theory where one can actually calculate a big portion of the relevant 
physics using the perturbation theory and renormalization-group 
techniques~\cite{som15,man00,neu94}.
 
Calculations of the cross sections of the ${\bar \Lambda}_c^- 
\Lambda_c^+$ production in the ${\bar p} p$ collisions have been 
reported in several publications using a variety of models 
\cite{kro89,kai94,tit08,gor09,hai10,hai11,kho12,shy14}. They employ 
varying degrees of freedom ranging from quarks
\cite{kro89,kai94,tit08,gor09,kho12} to meson baryon 
\cite{hai10,hai11,shy14} in the description of this reaction. However, 
the magnitudes of the predicted cross sections are strongly model 
dependent. On the other hand, calculations for other charmed-baryon 
channels have been reported only by a few authors. In Ref.~\cite{kho12}, 
total cross sections have been given for ${\bar \Sigma}_c^- \Lambda_c^+
$ and ${\bar \Sigma}_c^ -\Sigma_c^+$ channels, which are obtained by 
integrating the differential cross sections $d\sigma/dt$  ($t$ is the 
momentum transfer) over a limited range of $t$. In ref.~\cite{tit08}, 
$d\sigma/dt$ are provided for these final states, but the integrated 
cross sections are not given.

In Ref.~\cite{hai17}, cross sections have been presented for the 
charm-production channels  ${\bar p} p \to {\bar \Lambda}_c^- 
\Sigma_c^+$, ${\bar \Sigma}_c^- \Sigma_c^+$, ${\bar \Sigma}_c^0 
\Sigma_c^0$, ${\bar \Sigma}_c^{--} \Sigma_c^{++}$, ${\bar \Xi}_c^- 
\Xi_c^+$ and ${\bar \Xi}_c^0 \Xi_c^0$ within the J\"ulich 
meson-exchange model that was employed earlier~\cite{hai92a,hai92b} 
to investigate the ${\bar p} p \to {\bar \Lambda} \Lambda$ reaction. 
In this model, a coupled-channels framework is used that allows to 
take into account the initial- and final-state interactions in a 
rigorous way. The reaction proceeds via exchanges of appropriate 
mesons between ${\bar p}$ and $p$ leading to the final 
antibaryon-baryon states. For calculating the cross sections for the 
${\bar \Sigma}_c^0 \Sigma_c^0$, ${\bar \Sigma}_c^{--} \Sigma_c^{++}$, 
${\bar \Xi}_c^- \Xi_c^+$ and ${\bar \Xi}_c^0 \Xi_c^0$ final states, 
two-step mechanisms have been invoked in Ref.~\cite{hai17}. However,
in this reference, cross sections are reported only for beam 
momenta very close to the respective threshold of each final state.

In Ref.~\cite{shy14}, the ${\bar p} p \to {\bar \Lambda}_c^- 
\Lambda_c^+$ reaction has been studied within a single-channel 
effective Lagrangian model (see, e.g., Refs.~\cite{shy99,shy02}), 
where this reaction is described as a sum of the $t$-channel $D^0$ 
and $D^{*0}$ meson-exchange diagrams. The $s$- and $u$-channel 
resonance excitation terms are suppressed, as no resonance is known 
with energy in excess of 3.0 GeV having branching ratios for decays 
to the charmed-baryon channels. Furthermore, the direct ${\bar p}p$ 
annihilation into charmed-baryon final states via the contact 
diagrams is also suppressed due to the Okubo-Zweig-Iizuka condition. 

The aim of this paper is to extend the model of Ref.~\cite{shy14} to 
calculate the cross sections of the ${\bar p} p \to {\bar \Lambda}_c^- 
\Sigma_c^+$ and ${\bar p} p \to {\bar \Sigma}_c^- \Sigma_c^+$ 
reactions.  We provide predictions for the cross sections for beam 
momenta (${\bar p}_{lab}$) ranging from threshold to 18 GeV/$c$ in 
each case. Therefore, the range of ${\bar p}_{lab}$ of interest to 
the ${\bar P}ANDA$ experiment is well covered. As in Ref.~\cite{shy14}, 
these reactions are described as a sum of $t$-channel $D^0$ and 
$D^{*0}$ meson-exchange diagrams. The $s$- and  $u$-channel resonance 
excitation diagrams are suppressed due to the same reason as stated 
in the previous paragraph. Of course, like the  ${\bar \Lambda}_c^- 
\Lambda_c^+$ channel,  experimental data also do not exist at present  
for the other charmed-baryon channels. 

The amplitudes calculated in the effective Lagrangian model include 
basically only the Born terms. However, from the studies of the 
${\bar p} p \to {\bar \Lambda}_c^- \Lambda_c^+$ reaction, it is 
known that there is a strong sensitivity of the cross sections to 
the distortion effects in the initial and final states
\cite{gen84,tab85,kro87,bur88,koh86,bri89,rob91,tab91,hai92a,hai92b,alb93}.
We account for such effects approximately by using an eikonal 
approximation-based procedure.
  
In the next section, we present our formalism. The results and 
discussions of our work are given in Sec. III. Finally, the summary and 
conclusions of this study are presented in Sec. IV. 
 
\section{Formalism}

\begin{figure}[t]
\centering
\includegraphics[width=.50\textwidth]{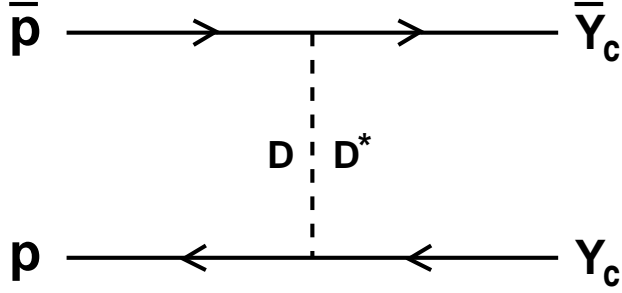}
\caption{
Graphical representation of the model used to describe the ${\bar p} + p 
\to {\bar Y}_c + Y_c$ reaction, where $Y_c$ represents a charmedr baryon. 
$D$ and $D^*$, in the intermediate line, represent the exchanges of $D$ 
pseudoscalar and $D^*$ vector mesons, respectively. In cases of ${\bar 
\Lambda}_c^- \Lambda_c^+$, ${\bar \Lambda}_c^- \Sigma_c^+$, and ${\bar 
\Sigma}_c^- \Sigma_c^+$ final states $D$ and $D^*$ correspond to  $D^0$ 
and $D^{*0}$ mesons, respectively.
}
\label{fig:Fig1}
\end{figure}

To evaluate various amplitudes for the processes shown in Fig.~1, we have 
used the effective Lagrangians at the charmed-baryon-meson-nucleon 
vertices, which are taken from Refs.~\cite{hob13,hai11a,gro90}. For the 
pseudoscalar $D$ meson exchange vertices, we have 
\begin{eqnarray}
{\cal L}_{BDN} & = & ig_{BDN}{\bar \psi}_B i\gamma_5 \psi_N \phi_{D} + H.c.,
\label{Eq.1}
\end{eqnarray}
where ${\psi}_B$ and $\psi_N$ are the charmed-baryon and nucleon 
(antinucleon) fields, respectively. In Eq.~(1), $\phi_{D}$ is the $D$ meson 
field and  $g_{BDN}$ represents the vertex coupling constant.

For the vector meson $D^{*}$ exchange vertices, the effective Lagrangian is
\begin{eqnarray}
{\cal L}_{BD^{*}N} & = & g_{BD^{*}N} {\bar \psi}_B \gamma_\mu \psi_N 
                         \theta_{D_{*}}^\mu + \frac{f_{BD^{*}N}}{4M}{\bar \psi}_B
                         \sigma_{\mu \nu} \psi_N F_{D^{*}}^{\mu \nu} + H.c.,
\label{Eq.2}
\end{eqnarray}
where $\theta_{D_{*}}^\mu$ is the vector meson field, with field strength tensor
$F_{D^{*}}^{\mu \nu} = \partial^\mu \theta_{D^{*}}^\nu - \partial^\nu 
\theta_{D^{*}}^\mu$. In Eq.~(2), $\sigma_{\mu \nu}$ is the usual tensor operator. 
The vector and tensor couplings are defined by $g$ and $f$, respectively. Their 
values at various vertices were adopted from Refs.~\cite{hob13,hai11a} as shown 
in Table I. The same couplings were used for the vertices involving both the 
proton and the antiproton. It was shown in Ref.~\cite{shy14} that the exchange 
of $D^{*0}$ dominates the $\Lambda_c^+ {\bar \Lambda}_c^-$ production reaction 
in the ${\bar p}-p$ collisions even for beam momenta closer to the production 
threshold. 

\begin{table}[t]
\begin{center}
\caption {Coupling constants at the $BD^0N$ and $BD^{*0}N$ vertices. These are 
taken from Refs.~\cite{hob13,hai11a}.  
}
\vspace{0.5cm}
\begin{tabular}{|c|c|c|c|c|}
\hline
Vertex  & Exchanged meson mass & Exchanged meson width & $g_{DBN}/\sqrt{4\pi}$ & 
$f_{DBN}/\sqrt{4\pi}$ \\
        &(MeV)                 & (MeV)                 &                       & 
                      \\
\hline
$ND^0\Lambda_c^+$      & 1864.84 & $-$  & 3.943  & $...$    \\
$ND^{*0}\Lambda_c^+$   & 2006.96 & 2.1   & 1.590 & 5.183   \\
$ND^0\Sigma_c^+$       & 1864.84 & $-$  & 0.759  & $...$    \\
$ND^{*0}\Sigma_c^+$    & 2006.96 & 2.1   & 0.918 & -2.222  \\
\hline
\end{tabular}
\end{center}
\end{table}

The coupling constants adopted by us at various vertices involved in the 
$t$-channel diagrams, were determined in Refs.~\cite{hai11a,hai07,hai08}, 
by using the SU(4) symmetry arguments in the description of the exclusive 
charmed-hadron production in the ${\bar D}N$ and $DN$ scattering within a 
one-boson-exchange picture. While, we acknowledge that the SU(4) symmetry 
will not hold rigorously, the deviations from the SU(4) coupling 
constants in the charm sector have been reported to be highly model 
dependent~\cite{kre12}. Recent calculations within light-cone sum rules 
suggest that deviations from the SU(4) values of the relevant coupling 
constants are not more than factors of 2~\cite{kho12}. On the other hand, 
in the constituent quark model calculations using the $^3P_0$ quark-pair 
creation mechanism, the deviations are at the most of the order of 
30$\%$~\cite{kre14}.

The off-shell behavior of the vertices is regulated by introducing form 
factors. Without them calculations with Born terms strongly overestimate 
the cross section at higher energies. Therefore, such contributions will 
have to be quenched with form factors. Another motivation for introducing 
form factors is that at higher energies one may expect sensitivity to the 
underlying quark structure of the hadrons. Because this physics is not 
included explicitly in our model, we can only account for it by 
introducing the phenomenological form factors at the vertices. In our 
approach, the form factors are not known \emph{a priori} and thus they 
introduce a certain arbitrariness in the calculations. In the current 
paper we limit ourselves to dipole form factors (see, e.g., Refs.
\cite{shy99,shy02}) at the vertices involving the pseudoscalar $D$ 
meson exchange because of their simplicity:
\begin{eqnarray}
F_i(q_i) & = & \frac{\lambda_i^2-m_{D_i}^2}{\lambda_i^2-q_{D_i}^2},
\label{Eq.3}
\end{eqnarray}
where $q_{D_i}$ is the momentum of the {\it i}th exchanged meson with 
mass $m_{D_i}$. $\lambda_i$ is the corresponding cutoff parameter, which 
governs the range of suppression of the contributions of high momenta 
carried out via the form factor. We chose a value of 3.0 GeV for 
$\lambda_i$ at both the vertices. The same $\lambda_i$ was also used in 
the monopole form factor employed in the studies presented in Refs.
\cite{hai10,hai11,hob13}. 

However, at the vertices involving the exchange of the vector meson 
$D^*$, we have used a different functional form of the form factor
\begin{eqnarray}
F_{i}(q_i) & = & \left [\frac{{\lambda_i^4}} {{\lambda_i^4} + 
(q_i^2-m_i^2)^2} 
\right ],
\label{Eq.4}
\end{eqnarray}
The argumentation for this different choice is presented in the 
discussion of the $\Sigma$-photoproduction results in Ref.~\cite{uso05}. 
Often, different functional forms and cutoff values are introduced for 
$t$-channel form factors~\cite{shy08,hai14,shy16a}. Although this can 
easily be motivated, it introduces additional model dependence and 
increases the number of free parameters. To limit the overall number of 
parameters we have taken the form factor given by Eq.~3 with a cutoff 
parameter $\lambda_i$, of 3.0 GeV for all the graphs involving the $D$ 
meson exchange, and that given by Eq.~4, with a $\lambda_i$ of 2.7 GeV 
for all the terms involving the $D^*$ meson exchange. Because the 
experimental data are not yet available for the reactions under 
investigation in this paper, it is not possible to put a more definite 
constraint on these quantities. Therefore, we restrict ourselves to 
these choices of the form factors and the $\lambda_i$ values.  

For calculating the amplitudes, we require the propagators for the 
exchanged mesons.  For the $D$ and $D^{*}$ mesons, the propagators are 
given by  
\begin{eqnarray}
G_{D}(q) & = & \frac{i} {{q^2 - m_{D}^2 }}, \label{Eq.5}\\
G_{D^{*}}^{\mu\nu}(q) & = & -i\left(\frac{{g^{\mu\nu}-q^\mu q^\nu/q^2}}
                      {{q^2 - (m_{D^{*}}-i\Gamma_{D^{*}}/2)^2}} \right).
\label{Eq.6}
\end{eqnarray}
In Eq.~(6), $\Gamma_{D^{*}}$ is the total width of the $D^{*}$ meson, which 
is given in table I taken from the latest Particle Data Group estimates
\cite{pat16}. 

After having established the effective Lagrangians, coupling constants, and 
forms of the propagators, the amplitudes of various diagrams can be written 
by following the well-known Feynman rules. The signs of these amplitudes are 
fixed by those of the effective Lagrangians, the coupling constants, and the 
propagators as described above. These signs are not allowed to change 
anywhere in the calculations.
 
Next, we describe how the initial- and final-state interactions are taken 
into account in our calculations. We note that, for the ${\bar p} p$ initial 
state, the annihilation channel is almost as strong as the elastic scattering
channel. This large depletion of the flux can be accounted for by introducing 
absorptive potentials that are used in optical models or in coupled-channels 
approaches~\cite{koh86,hai92a,hai92b,alb93,hai10}. In this work, we do not 
employ such a detailed treatment. Instead, we use a procedure that was 
originated by Sopkovich~\cite{sop62}. In this method, the transition amplitude 
with distortion effects is written as
\begin{eqnarray}
T^{{\bar p}p \to {\bar Y}_c Y_c} & = &
\sqrt{\Omega^{{\bar p}p}}T^{{\bar p}p \to {\bar Y}_c Y_c}_{Born}
\sqrt{\Omega^{{\bar Y}_c Y_c}} 
\label{Eq.7}
\end{eqnarray}
where $T^{{\bar p}p \to {\bar Y}_c Y_c}_{Born}$ is the transition matrix 
calculated within the Born approximation, and $\Omega^{{\bar p}p}$ and the 
$\Omega^{{\bar Y}_c Y_c}$ are the matrices describing the initial- and 
final-state elastic scatterings, respectively. Their effects are to dampen 
the wave functions and hence the amplitudes. For the case of the ${\bar p}p 
\to {\bar \Lambda}_c^- \Lambda_c^+$ reaction, it was shown in Ref.
\cite{hai10}, that because of the strong absorption in the initial channel, 
the results turned out to be  rather insensitive to the final-state $ {\bar 
\Lambda}_c^- \Lambda_c^+$ interactions. In fact, even if the final-state 
interactions (FSIs) were ignored totally, the total cross sections did not 
change by more than 10$\%$--15$\%$. We assume this to be true also for the 
other charmed-baryon final states as well. Therefore, in order to keep the 
number of free parameters small, we decided to fully neglect FSIs and 
concentrate only on the initial-state interaction in our calculations of 
the cross sections of all the charmed-baryon-antibaryon final channels.

For the present purpose, we neglect the real part of the proton-antiproton  
interaction and describe the strong absorption by an imaginary potential of 
Gaussian shape with range parameter $\mu$ and strength $V_0$
\begin{eqnarray}
U(b,z)& = & V_0 \,\,exp(-\mu^2 r^2),
\label{Eq.8}
\end{eqnarray}
where $\mu$ is the range parameter and $V_0$ the strength of the potential. 
In Eq.~8, $r^2 = b^2 + z^2$, with $b$ being the impact parameter of the 
collision. By using the eikonal approximation, the corresponding 
attenuation integral can be evaluated in a closed form. Similar to Refs.
\cite{sop62,rob91}, we obtain for $\Omega^{{\bar p}p}$  
\begin{eqnarray}
\Omega^{{\bar p}p} & = & exp\big[\frac{-\sqrt{\pi}EV_0}{\mu k}\,\, 
exp(-\mu^2 b^2)\big], \label{eq.9}
\end{eqnarray} 
where $E$ and $k$ are the center of mass energy and momentum of the 
particular channel, respectively. In our numerical calculations, we have 
used $V_0 = 0.8965$ GeV and $\mu = 0.3369$ GeV.  For the impact parameter, 
we have taken a value 0.327 GeV$^{-1}$. These parameters are the same as 
those used in Refs.~\cite{shy14,shy16a}. As shown in Ref.~\cite{shy14}, 
with these parameters, it was possible to get cross sections for the 
${\bar \Lambda}_c^- \Lambda_c^+$ production in close agreement with those 
calculated within the coupled-channels approach of Ref.~\cite{hai10}, where 
distortion effects are rigorously treated. 

Although the parameters $V_0$ and $\mu$ may change with energy, we have made 
them global; that is, they remain the same at all the energies corresponding to 
all the final channels. Thus, we have only three fixed parameters in our 
calculations of the initial-state distortion effects.

\section{Results and Discussions}

\begin{figure}[t]
\centering
\includegraphics[width=.50\textwidth]{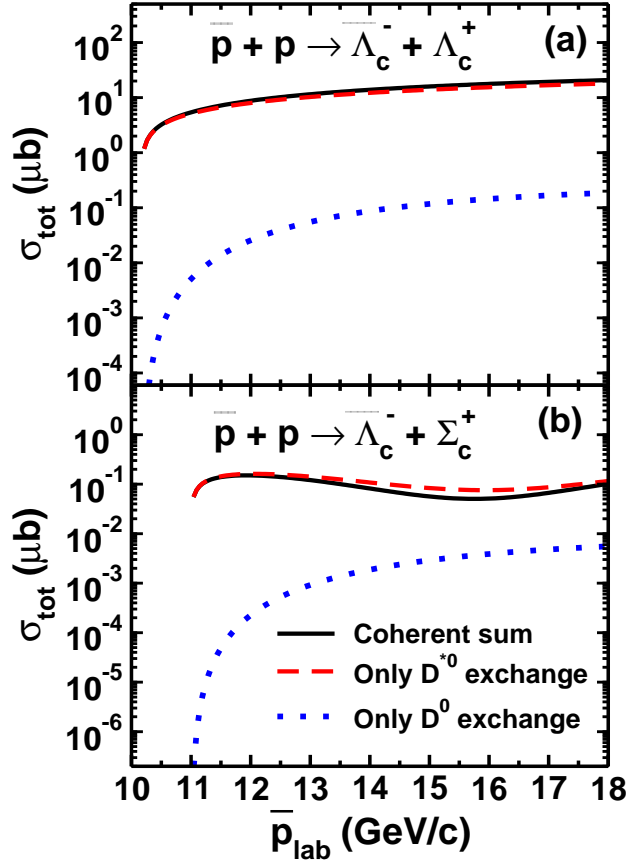}
\caption{(color online)
Total cross section for ${\bar p}p \to {\bar \Lambda}_c^- \Lambda_c^+$ (a) 
and ${\bar p}p \to {\bar \Lambda}_c^- \Sigma_c^+$ (b) reactions as a 
function of the beam momentum. It is to be noted that cross sections for the 
reaction ${\bar p}p \to {\bar \Sigma}_c^- \Lambda_c^+$ are the same as those 
shown in Fig.~2(b). In these figures, the contributions of $D^0$ and $D^{*0}$ 
exchange processes are shown by dotted and dashed lines, respectively. The 
solid line represents their coherent sum.
}
\label{fig:Fig2}
\end{figure}

In Fig. 2, we investigate the role of various meson-exchange processes in the 
total cross sections ($\sigma_{tot}$) of the reactions ${\bar p}p \to {\bar 
\Lambda}_c^- \Lambda_c^+$ [Fig.~2(a)], and ${\bar p}p \to {\bar \Lambda}_c^- 
\Sigma_c^+$ [Fig. 2(b)] as a function of ${\bar p}$ beam momenta. Although the 
predictions for the cross sections of the ${\bar p}p \to {\bar \Lambda}_c^- 
\Lambda_c^+$ reaction were presented already in Ref.~\cite{shy14}, we give it 
here again for the purpose of comparison with the $\sigma_{tot}$ of other 
charmed-baryon channels. Also, because in this paper we have used a different 
form factor at the vertices involving $D^{*0}$ meson exchange, it would be of 
interest to have results also for this charmed-baryon final state. In 
Figs.~2(a) and 2(b), total cross sections $\sigma_{tot}$ are shown for 
${\bar p}_{lab}$ varying in the range of threshold to 18 GeV/$c$ that covers 
the beam momenta of interest to the ${\bar P}ANDA$ experiment. We note that, 
for both the reactions, the cross sections increase gradually as 
${\bar p}_{lab}$ goes above the thresholds of the respective reactions. The 
threshold beam momenta for  ${\bar p}p \to {\bar \Lambda}_c^- \Lambda_c^+$ 
and ${\bar p}p \to {\bar \Lambda}_c^- \Sigma_c^+$ reactions are 10.162  and 
10.99 GeV/$c$, respectively. For ${\bar p}_{lab}$ around 15 GeV/$c$, which 
is the beam momentum region of interest to the ${\bar P}ANDA$ experiment, 
$\sigma_{tot}$ for the ${\bar p}p \to {\bar \Lambda}_c^- \Lambda_c^+$ 
reaction is about one order of magnitude larger than that for the ${\bar p}p 
\to {\bar \Lambda}_c^- \Sigma_c^+$ reaction. The likely reasons for this 
difference are the smaller coupling constants at the $ND^{*} \Sigma_c^+$ 
vertices and the negative interference between the $D^{*0}$ and $D^0$ 
exchange terms in case of the ${\bar \Lambda}_c^- \Sigma_c^+$ final state.
 
In Figs.~2(a) and 2(b), we note that the $D^{*0}$ exchange process dominates 
the cross sections for both the final states. The  $D^0$ exchange contributions 
are nearly 2 orders of magnitude smaller than those of the $D^{*0}$ exchange 
in case of the ${\bar \Lambda}_c^- \Lambda_c^+$ final state and nearly an order 
of magnitude for the ${\bar \Lambda}_c^- \Sigma_c^+$ final state in the region 
of  higher beam momenta. Interestingly, we notice in Fig~2(b) that, even though 
for ${\bar p}_{lab}$ beyond 14 GeV/$c$ the individual contributions of the 
$D^0$ exchange terms are at least 1 order of magnitude smaller, they still 
influence the total cross sections significantly through the interference terms 
that are destructive in this case.
\begin{figure}[t]
\centering
\includegraphics[width=.50\textwidth]{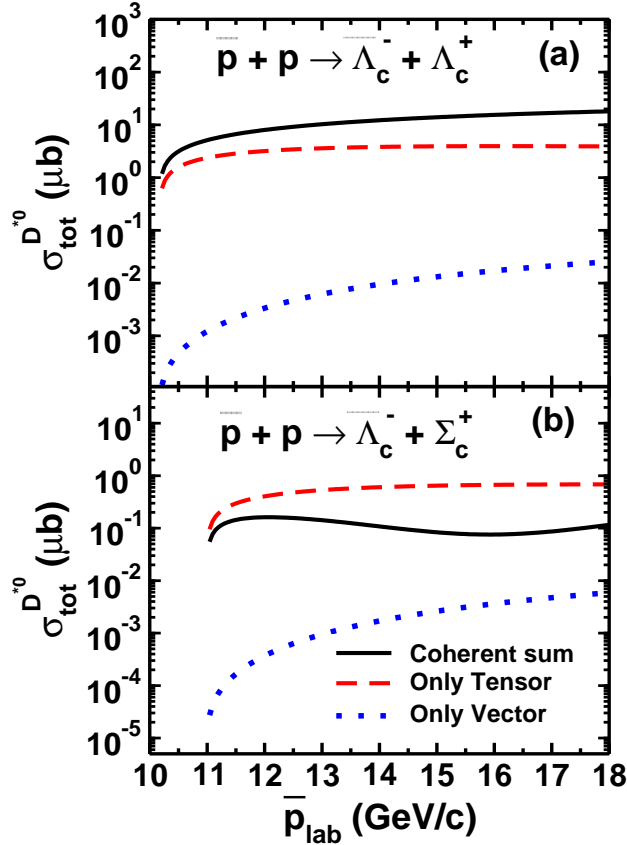}
\caption{(color online)
Contributions of the vector and tensor coupling terms to the $D^{*0}$ 
meson-exchange component of the total cross section ($\sigma_{tot}^{D^{*0}}$) 
for reactions  ${\bar p}p \to {\bar \Lambda}_c^- \Lambda_c^+$ (a)
and ${\bar p}p \to {\bar \Lambda}_c^- \Sigma_c^+$ (b) as a function of the 
antiproton beam momentum. The contributions of the tensor and vector terms are 
shown by the dashed and dotted lines, respectively.  The solid line represents 
their coherent sum.
}
\label{fig:Fig3}
\end{figure}

The domination of the $D^{*0}$ meson-exchange terms is related to the presence 
of the strong tensor part in the $ND^{0*}Y_c^+$ couplings. We note from Table 
I that the ratio of tensor to vector coupling constants of these vertices 
is in the vicinity of three. Also there is  additional momentum dependence induced 
by the derivative coupling in the tensor interaction part of the effective 
Lagrangian. In Figs.~3(a) and (3b), we show the individual contributions of 
tensor and vector terms to the $D^{*0}$ meson-exchange component of the total 
cross section ($\sigma_{tot}^{D^{*0}}$) for the reactions  ${\bar p}p \to {\bar 
\Lambda}_c^- \Lambda_c^+$, and  ${\bar p}p \to {\bar \Lambda}_c^- \Sigma_c^+$,
respectively. It is clear that the tensor coupling terms make the dominant 
contributions to $\sigma_{tot}^{D^{*0}}$ for both the reactions. Although the 
vector coupling terms are relatively quite small in themselves, they can   
influence the cross sections $\sigma_{tot}^{D^{*0}}$ significantly through the
interference terms. For  the ${\bar p}p \to {\bar \Lambda}_c^- \Lambda_c^+$ 
reaction the interference of tensor and vector terms is constructive that 
enhances $\sigma_{tot}^{D^{*0}}$ particularly at higher beam momenta. However 
for the ${\bar p}p \to {\bar \Lambda}_c^- \Sigma_c^+$ reaction, this 
interference is destructive that reduces $\sigma_{tot}^{D^{*0}}$ rather sharply 
in this region. 
\begin{figure}[t]
\centering
\includegraphics[width=.50\textwidth]{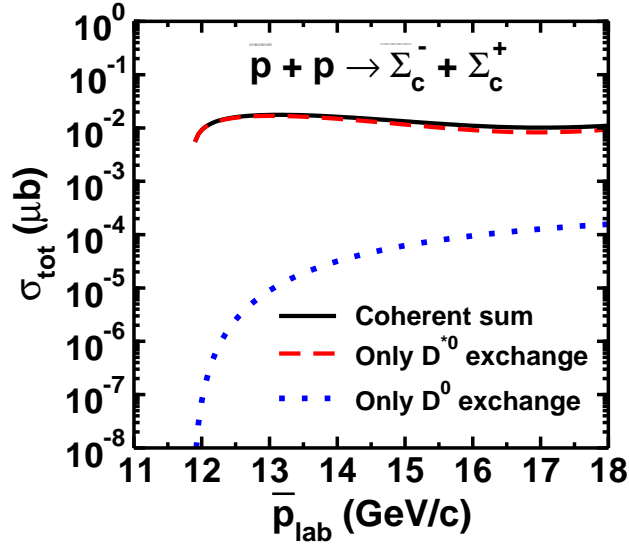}
\caption{(color online)
Contributions of the $D^0$ (dotted line) and $D^{*0}$ (dashed line) meson 
exchange processes to the total cross section of the ${\bar p}p \to {\bar 
\Sigma}_c^- \Sigma_c^+$ reaction (solid line) as a function of the antiproton 
beam momentum. 
}
\label{fig:Fig4}
\end{figure}

It would be of interest to compare our cross sections for these two channels 
to those published previously. In Ref.~\cite{hai17}, the cross section 
$\sigma_{tot}$ are given for the ${\bar \Lambda}_c^- \Lambda_c^+$ channel 
for ${\bar p}_{lab}$ in the range of threshold to 11.5 GeV/$c$. In this range 
of beam momenta our $\sigma_{tot}$ for this reaction are approximately in 
agreement with those calculated within the meson-exchange model (MEM) in Ref.
\cite{hai17}. On the other hand, for the ${\bar \Lambda}_c^- \Sigma_c^+$ 
channel our cross sections are smaller than the MEM cross sections of Ref.
\cite{hai17} by factors of about 10. In Ref.~\cite{kho12}, $\sigma_{tot}$ 
for the ${\bar \Lambda}_c^- \Lambda_c^+$ and ${\bar \Lambda}_c^- \Sigma_c^+$ 
channels are given around the ${\bar p}_{lab}$ of 15 GeV. These authors have 
performed their calculations within a nonperturbative quark-gluon string 
model~\cite{kai94} employing the baryon-meson coupling constants from 
light-cone sum rules. Our cross sections are at least an order of magnitude 
larger than those of Ref.~\cite{kho12} for both the channels. Similarly, the 
cross sections for the ${\bar \Lambda}_c^- \Lambda_c^+$ production channel 
reported in Refs.~\cite{kai94,tit08,gor09} are also significantly smaller 
than those predicted in our study. Thus, differences between our cross 
sections and those of Refs.~\cite{kho12,kai94,tit08,gor09} are substantial 
for beam momenta relevant to the ${\bar P}ANDA$ experiment.  
     
In Fig.~4, we investigate the relative contributions of $D^0$ and $D^{*0}$ 
meson-exchange processes to the total cross section $\sigma_{tot}$ of the 
${\bar p}p \to {\bar \Sigma}_c^- \Sigma_c^+$ reaction. We first notice that 
the magnitudes of the $\sigma_{tot}$ are further reduced in this case as 
compared to those of the ${\bar \Lambda}_c^- \Sigma_c^+$ production. This is 
due to the fact that now there are two $ND^{*} \Sigma_c^+$ vertices having 
coupling constants smaller than that of the $ND^{0*} \Lambda_c^+$ vertex. 
Like the results presented in Fig~2, in this case also the $D^{*0}$ exchange 
terms make the predominant contributions to $\sigma_{tot}$. At the beam 
momenta around 15 GeV/$c$, the $D^0$ meson-exchange terms are about two 
order magnitudes smaller than those of the $D^{0*}$ exchange. At beam momenta 
closer to the threshold (${\bar p}_{lab}$ = 11.85 GeV/$c$) the differences 
between the contributions of two meson-exchange terms are even bigger. 

In Fig.~5, the relative contributions of the vector and tensor terms of the 
$ND^{0*}\Sigma_c^+$ couplings to cross section $\sigma_{tot}^{D^{*0}}$ are 
shown. We see that the tensor component of the coupling dominates 
$\sigma_{tot}^{D^{*0}}$. It is further noted that, like the results shown in 
Fig.~3(b), the interference between tensor and vector amplitudes is 
destructive. This effect significantly reduces the magnitude of 
$\sigma_{tot}^{D^{*0}}$ at higher beam momenta. 

It may be possible to calculate the cross sections of the ${\bar p}p \to {\bar 
\Sigma}_c^0 \Sigma_c^0$ and ${\bar p}p \to {\bar \Sigma}_c^{--} \Sigma_c^{++}$ 
channels by also invoking some two-step mechanism as is done in Ref.
\cite{hai17}.  However, it may not be feasible in the version of the effective 
Lagrangian model presented in this paper. Nevertheless, the magnitudes of the 
cross sections for these channels are similar to those of the 
${\bar \Sigma}_c^{-} \Sigma_c^{+}$ channel as is shown in Ref.~\cite{hai17}. 
\begin{figure}[t]
\centering
\includegraphics[width=.50\textwidth]{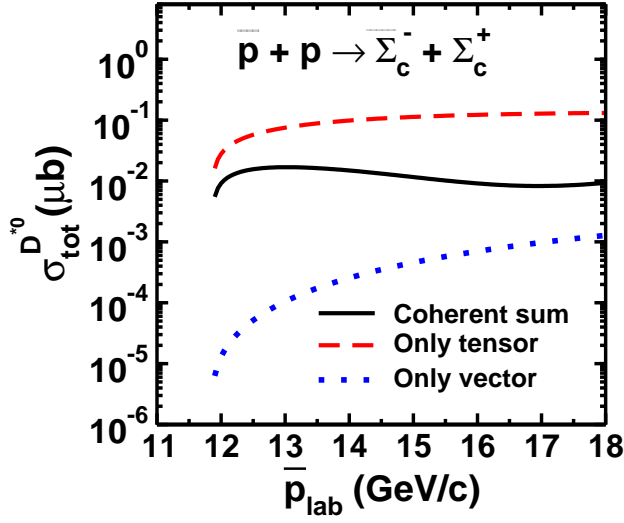}
\caption{(color online)
Contributions of the vector and tensor coupling terms to the $D^{*0}$ 
exchange component of the cross sections ($\sigma_{tot}^{D^{*0}}$) for the 
${\bar p}p \to {\bar \Sigma}_c^- \Sigma_c^+$ reaction as a function of the 
beam momentum ${\bar p}_{lab}$. The tensor and vector terms are shown by the 
dashed and dotted lines, respectively. The solid line represents their 
coherent sum.
}
\label{fig:Fig5}
\end{figure}

\section{Summary and conclusions}

In summary,  we studied the ${\bar p}p \to {\bar \Lambda}_c^- 
\Lambda_c^ + $, ${\bar p}p \to {\bar \Lambda}_c^- \Sigma_c^+$, ${\bar p}p 
\to {\bar \Sigma}_c^- \Lambda_c^+$, and ${\bar p}p \to {\bar \Sigma}_c^-
\Sigma_c^+ $ reactions using a phenomenological effective Lagrangian model 
that involves the meson-baryon degrees of freedom. The charmed-baryon 
production mechanism is described by the $t$-channel $D^0$ and $D^{*0}$ 
meson-exchange diagrams, while largely phenomenological initial- and 
final-state interactions were used to account for the distortion 
effects. The coupling constants at various vertices have been taken from 
the $DN$ and ${\bar D}N$ scattering studies reported in Refs.
\cite{hai07,hai08,hai11a}. The off-shell corrections at the $D^0$ meson 
vertices are incorporated by introducing the monopole form factor, which 
was taken to be the same for all the cases. However, at vertices 
involving the $D^{*0}$ meson exchange, a type of quadrupole form factor 
has been used. The shapes of the form factors and the values of the 
cutoff parameters appearing therein have been held fixed for all the 
final charmed-baryon production channels.

In the range of beam momenta of interest to the ${\bar P}ANDA$ experiment, 
the total cross sections are largest for the  ${\bar \Lambda}_c^- \Lambda_c^
+ $ production channel. For ${\bar p}_{lab}$ not too far from the 
threshold, the cross sections for this final state as predicted by our 
model are similar to those obtained within the J\"ulich meson-exchange 
model as reported in Ref.~\cite{hai17}. The production cross sections for 
the ${\bar \Lambda}_c^- \Sigma_c^+$ and ${\bar \Sigma}_c^- \Sigma_c^+$ final 
states are smaller than those of the ${\bar \Lambda}_c^- \Lambda_c^+ $ 
channel by factors of the order of 10 and 100, respectively. The reasons
for this is traced back to the large negative interference between the 
vector and tensor parts of the $D^{*0}$ meson-exchange term and relatively 
smaller coupling constants of the $ND^{*0}\Sigma_c^+$ vertices. Furthermore, 
even though our cross sections for these channels are smaller than those 
calculated within the meson-exchange model of Ref.~\cite{hai17}, they are 
still larger than those of Refs.~\cite{kai94,tit08,gor09,kho12}. Because 
these earlier calculations have used different types of approaches, which 
employ quark-model-based ingredients in their calculations, it is not 
trivial to locate the reason for the large difference seen between their 
and our results. The ${\bar P}ANDA$ experiment for these reactions should 
provide an opportunity to understand this difference once the FAIR facility 
is operational. If the cross sections are as large as predicted in our 
calculations as well as in those of the meson-exchange model of Ref.
\cite{hai17}, the experimental requirements may perhaps become less 
stringent for their measurements at ${\bar P}ANDA$.

We found that the vector meson ($D^{*0}$) exchange terms dominate the cross 
sections for all the reaction channels in the entire range of beam momenta. 
The reasons for the large strength of this exchange process are the strong 
tensor coupling of the vector mesons (similar to the large tensor coupling 
of the $\rho$ meson in $NN$ interactions), and the additional momentum 
dependence introduced by the derivative part of the corresponding interaction. 
Although, the individual contributions of the $D^0$ exchange terms are 
relatively weak, they can still affect the total cross sections through the 
interference terms.

We treated the initial- and final-state interactions within an eikonal 
approximation-based phenomenological method. Generally, the parameters of 
this model are constrained by fitting to the experimental data. Because of 
the lack of any experimental information, it has not been possible to test 
our model thoroughly. Therefore, there may be some uncertainty in the 
absolute magnitudes of our cross sections. Nevertheless, our near-threshold 
cross sections for the ${\bar \Lambda}_c^ - \Lambda_c^+ $ production channel 
are very close to those of Refs.~\cite{hai17}, where distortion effects have 
been treated more rigorously within a coupled-channels approach. There may 
also be some uncertainty in our cross sections coming from the shapes of the 
form factors and the values of the cutoff parameter involved therein. In 
models like ours choice for these quantities is guided by their ability to
reproduce the experimental data~\cite{shy10}, which is not feasible at this 
stage for the charmed-baryon production. We have tried to minimize the 
effects of such uncertainties to some extent by using the same shape of the 
form factor and the same value of the cutoff parameter in calculations of 
all the final charmed-baryon channels. Hopefully, concrete experiments that 
are likely to be pursued with the ${\bar P}ANDA$ experiment at FAIR will help 
in removing most of these uncertainties.  
   
\section{acknowledgments}
 
This work has been supported by the Science and Engineering Research Board (SERB), 
Department of Science and Technology, Government of India under Grant No. 
SB/S2/HEP-024/2013.


\begin{thebibliography}{99}

\bibitem{dub03}
M.S. Dubrovin (CLEO Collaboration), hep-ex/0305006.

\bibitem{zie11}
V. Ziegler ($BABAR$ Collaboration),  AIP Conf. Proc. {\bf 1374}, 577 (2011), and 
references therein.

\bibitem{kat14}
Y. Kato (Belle Collaboration), Proc. Sci., Hadron2013 (2014), 053, and references 
therein.

\bibitem{wie11}
U. Wiedner, Prog. Part. Nucl. Phys. {\bf 66}, 477 (2011).

\bibitem{ern09}
W. Erni {\it et al.}, arXiv:0903.3905 [hep-ex].

\bibitem{dov77}
C. B. Dover and S. H. Kahana, Phys. Rev. Lett. {\bf 39}, 1506 (1977).

\bibitem{shy17}
R. Shyam and K. Tsushima, Phys. Lett. B {\bf 770}, 236 (2017).

\bibitem{tsu99}
K. Tsushima, D. H. Lu, A. W. Thomas, K. Saito, and R. H. Landau,
Phys. Rev. C {\bf 59}, 2824 (1999).

\bibitem{gar10}
C. Garcia-Recio, J. Nieves, and L. Tolos, Phys. Lett. B {\bf 690}, 369 
(2010).

\bibitem{gar12}
C. Garcia-Recio, J. Nieves, L. L. Salcedo, and L. Tolos, Phys. Rev. C
{\bf 85}, 025203 (2012).

\bibitem{yam16}
J. Yamagata-sekihara, C. Garcia-Recio, J. Nieves, L. L. Salcedo, and
L. Tolos, Phys. Lett. B {\bf 754}, 26 (2016).

\bibitem{tsu03}
K. Tsushima and F. C. Khanna, Phys. Lett. B {\bf 552}, 138 (2003).

\bibitem{tsu04}
K. Tsushima and F. C. Khanna, J. Phys. G: Nucl. Part. Phys. {\bf 30}, 
1765 (2004).

\bibitem{jim11}
C. E. Jimenez-Tejero, L. Tolos, I. Vida\~na, and A. Ramos, Few-Body Syst.
{\bf 50}, 351 (2011).
 
\bibitem{tol13}
L. Tolos, Int. J. Mod. Phys. E, {\bf 22}, 1330027 (2013).
 
\bibitem{kre16}
G. Krein, AIP Conf. Proc. {\bf 1701}, 020012 (2016).
 
\bibitem{shy16}
R. Shyam and K. Tsushima, Phys. Rev. D {\bf 94}, 074041 (2016).

\bibitem{som15} 
Rainer Sommer, arXiv:1501.03060 [hep-lat]. 

\bibitem{man00}
A. V. Manohar and M. B. Wise, {\it Cambridge Monographs on Particle Physics, 
Nuclear Physics and Cosmology}, (Cambridge University Press, Cambridge, 
England, 2000), Vol. 10.

\bibitem{neu94}
M. Neubert, Phys. Rep. {\bf 245}, 259 (1994).

\bibitem{kro89}
P. Kroll, B. Quadder, and W. Schweiger, Nucl. Phys. {\bf B316}, 373 (1989).

\bibitem{kai94}
A. B. Kaidalov  and P. E. Volkovitsky, Z. Phys. C {\bf 63}, 517 (1994).

\bibitem{tit08}
A. I. Titov and B. K\"ampfer, Phys. Rev. C {\bf 78}, 025201 (2008).

\bibitem{gor09}
A. T. Goritschnig, P. Kroll, and W. Schweiger, Eur. Phys. J. A {\bf 42}, 
43 (2009). 

\bibitem{hai10}
J. Haidenbauer, and G. Krein, Phys. Lett. B {\bf 678}, 314 (2010).

\bibitem{hai11}
J. Haidenbauer, and G. Krein, Few-Body Syst. {\bf 50}, 183 (2011).

\bibitem{kho12}
A. Khodjamirian, Ch. Krein, Th. Mannel, and Y.-M. Wang, Eur. Phys. J. A 
{\bf 48}, 31 (2012).

\bibitem{shy14}
R. Shyam and H. Lenske, Phys. Rev. D {\bf 90}, 014017 (2014).

\bibitem{hai17}
J. Haidenbauer and G. Krein, Phys. Rev. D {\bf 95}, 014017 (2017).

\bibitem{hai92a}
J. Haidenbauer, T. Hippchen, K. Holinde, B. Holzenkamp, V. Mull, and 
J. Speth, Phys. Rev. C {\bf 45}, 931 (1992).

\bibitem{hai92b}
J. Haidenbauer, K. Holinde, V. Mull and J. Speth, Phys. Rev. C {\bf 46}, 
2158 (1992).

\bibitem{shy99}
R. Shyam, Phys. Rev. C {\bf 60}, 055213 (1999).

\bibitem{shy02}
R. Shyam and U. Mosel, Phys. Rev. C {\bf 67}, 065202 (2003).

\bibitem{gen84}
H.~Genz and S.~Tatur, Phys. Rev. D {\bf 30}, 63 (1984).

\bibitem{tab85}
F.~Tabakin and R.~A.~Eisenstein, Phys. Rev. C {\bf 31}, 1857 (1985).

\bibitem{kro87}
P. Kroll and W. Schweiger, Nucl. Phys. {\bf A474}, 608 (1987).

\bibitem{bur88}
M.~Burkardt and M.~Dillig, Phys. Rev. C {\bf 37}, 1362 (1988).

\bibitem{koh86}
M. Kohono and W. Weise, Nucl. Phys. A {\bf 454}, 429 (1986).

\bibitem{bri89}
G.~Brix, H.~Genz, and S.~Tatur, Phys. Rev. D {\bf 39}, 2054 (1989).

\bibitem{rob91}
W. Roberts, Z. Phys. C {\bf 49}, 633 (1991).

\bibitem{tab91}
F.~Tabakin, R.~A.~Eisenstein, and Y. Lu, Phys. Rev. C {\bf 44}, 1749 (1991).

\bibitem{alb93}
M.~A.~Alberg, E.~M.~Henley, L. Wilets, P.~D. Kunz, Nucl. Phys. {\bf A560},
365 (1993).

\bibitem{hob13}
T.~J.~Hobbs, J.~T.~Londergan, and W. Melnitchouk, Phys. Rev. D 89, 074008 (2014) 

\bibitem{hai11a}
J. Haidenbauer, G. Krein, U.-G. Meissner, and L. Tolos, Eur. Phys. J. A 
{\bf 47}, 18 (2011).

\bibitem{gro90}
A. M\"uller-Groeling, K.~Holinde, and J.~Speth, Nucl. Phys. {\bf A513}, 557 (1990). 

\bibitem{hai07}
J. Haidenbauer, G. Krein, U.-G. Meissner, and A. Sibirtsev, Eur. Phys. J. A 
{\bf 33}, 107 (2007).

\bibitem{hai08}
J. Haidenbauer, G. Krein, U.-G. Meissner, and A. Sibirtsev, Eur. Phys. J. A 
{\bf 37}, 55 (2008).

\bibitem{kre12}
G. Krein, POS (Confinement X) (2012) 144.

\bibitem{kre14}
G. Krein, EPJ Web Conf. {\bf 73} (2014) 05001.

\bibitem{uso05}
A. Usov and O. Scholten, Phys. Rev. C{\bf 72}, 025205 (2005).

\bibitem{shy08}
R. Shyam and O. Scholten, Phys. Rev. C {\bf 78}, 065201 (2008).

\bibitem{hai14}
J. Haidenbauer and G. Krein, Phys. Rev. D {\bf 89}, 114003 (2014).

\bibitem{shy16a}
R. Shyam and H. Lenske, Phys. Rev. D {\bf 93}, 034016 (2016).

\bibitem{pat16}
C. Patrignani {\it et al.} (Particle Data Group), Chin. Phys. C, 40, 
100001 (2016) and 2017 update.

\bibitem{sop62}
N. J. Sopkovich, Nuovo Cimento {\bf 26}, 186 (1962).

\bibitem{shy10}
R. Shyam, O. Scholten and H. Lenske, Phys. Rev. C {\bf 81}, 015204 (2010).
\end{thebibliography}
\end{document}